\documentclass[letterpaper]{article}
\usepackage{times}
\usepackage{helvet}
\usepackage{courier}
\usepackage[hyphens]{url}
\usepackage{graphicx}
\usepackage{amsmath}
\usepackage{amssymb}
\usepackage{booktabs}
\usepackage{natbib}
\usepackage{caption}
\usepackage[utf8]{inputenc}
\usepackage{graphicx}
\usepackage{booktabs}
\usepackage{pdflscape}
\usepackage{adjustbox}
\usepackage{multicol}
\usepackage{amsmath,amssymb}
\usepackage{microtype} 
\usepackage[T1]{fontenc}
\usepackage{hyperref}
\usepackage{authblk}
\usepackage{ragged2e}
\usepackage{titlesec}
\usepackage{multirow}
\usepackage[none]{hyphenat}  
\usepackage{natbib} 
\usepackage{setspace} 
\usepackage{blindtext}
\usepackage[a4paper, total={7in, 10in}]{geometry}
\usepackage{hyperref} 
\usepackage[most]{tcolorbox}
\usepackage{geometry}
\usepackage{booktabs}
\usepackage{longtable}
\usepackage{caption}
\usepackage{cleveref}
\newtcolorbox{highlightblock}{
  breakable,              
  enhanced,
  colback=blue!40,       
  colframe=blue!40,      
  boxrule=0pt,             
  sharp corners,
  interior style={opacity=0.6}, 
    }

\hypersetup{
    colorlinks = true,      
    linkcolor  = blue,      
    citecolor  = magenta,   
    urlcolor   = red       
}

\date{\today\\[3em]}

\title{Role of Large Language Models and Retrieval-Augmented Generation for Accelerating Crystalline Material Discovery: A Systematic Review}

\author[1]{Agada Joseph Oche\thanks{Corresponding author: \texttt{joe88data1@gmail.com}, ORCID: https://orcid.org/0009-0000-9479-3715}}
\author[2]{Arpan  Biswas\thanks{Corresponding author: \texttt{abiswas5@utk.edu}, ORCID: https://orcid.org/0000-0002-4054-7700}}

\affil[1]{Bredesen Center for Interdisciplinary Research, University of Tennessee, Knoxville, USA, 37996}
\affil[2]{University of Tennessee-Oak Ridge Innovation Institute, University of Tennessee, Knoxville, USA, 37996}

\begin{document}
\maketitle

\begin{abstract}
Large language models (LLMs) have emerged as powerful tools for knowledge-intensive tasks across domains. In materials science, to find novel materials for various energy efficient devices for various real-world applications, requires several time and cost expensive simulations and experiments. In order to tune down the uncharted material search space, minimizing the experimental cost, LLMs can play a bigger role to first provide an accelerated search of promising known material candidates. Furthermore, the integration of LLMs with domain-specific information via retrieval-augmented generation (RAG) is poised to revolutionize how researchers predict materials structures, analyze defects, discover novel compounds, and extract knowledge from literature and databases. In motivation to the potentials of LLMs and RAG in accelerating material discovery, this paper presents a broad and systematic review to examine the recent advancements in applying LLMs and RAG to key materials science problems. We survey state-of-the-art developments in crystal structure prediction, defect analysis, materials discovery, literature mining, database integration, and multi-modal retrieval, highlighting how combining LLMs with external knowledge sources enables new capabilities. We discuss the performance, limitations, and implications of these approaches, and outline future directions for leveraging LLMs to accelerate materials research and discovery for advancement in technologies in the area of electronics, optics, biomedical, and energy storage.
\end{abstract}

\newpage
\begin{multicols}{2}

\section{Introduction}
The rapid progress of \textit{large language models} (LLMs) in natural language processing has begun to significantly impact scientific domains, including materials science \citep{Yu2024-GameChangers}. LLMs such as GPT-3 and GPT-4, built on the Transformer architecture \citep{Vaswani2017} and trained on massive text corpora \citep{Brown2020-GPT3}, demonstrate impressive abilities in generating human-like text and reasoning over knowledge. However, out-of-the-box LLMs often lack specialized scientific knowledge and can produce \textit{hallucinations} (incorrect statements) when tasked with domain-specific problems. This has led to growing interest in techniques to ground LLMs with reliable external information \citep{Jiang2025-NPJReview, Lewis2020-RAG}. In materials science, there is a vast region of material search space, suited for various energy efficient devices for various real-world applications. One of the major challenges is to find intended material candidates, which require several time and cost expensive simulations \cite{Barghathi:2022bw, CasianoDiaz:2023pi} and experiments \cite{Harris2024PLD}. With current global motivation of AI-driven accelerated search for material discovery, LLM can play a bigger role to search quickly into the known material space to find application-driven promising candidates. This accelerated guided search in the known material space, can potentially help to tune down the vast uncharted material space for faster discovery of novel materials. This motivates us to first look into the ongoing researches and potential of LLMs in accelerating material research, in the early phase of discovery. 

Early applications of natural language processing (NLP) focused on text mining and knowledge extraction from literature, for example using word embeddings to capture structure-property relations \citep{Tshitoyan2019}. LLMs promise a more flexible and powerful approach, able to interpret complex queries, generate hypotheses, and interface with databases or simulation tools through natural language. \textit{Retrieval-augmented generation} (RAG) has emerged as a key strategy for domain adaptation: an LLM is coupled with a retrieval module that supplies relevant documents or data from an external knowledge base \citep{Lewis2020-RAG}. By injecting up-to-date and context-specific information into the LLM's prompt, RAG can significantly improve factual accuracy and domain relevance of the generated answers. This review provides a systematic overview of how LLMs and RAG are being leveraged in materials science and engineering. We cover applications ranging from crystal structure prediction and defect analysis to materials discovery and literature mining, highlighting the current state-of-the-art, challenges, and charting the way forward.

\section{Methods}
\label{sec:methods}
As per our research contribution in this paper, we performed a systematic literature survey to identify relevant publications on LLM and RAG applications in materials science. Multiple databases (Web of Science, Scopus, Google Scholar) and preprint servers (arXiv) were queried using keywords such as “large language model”, “retrieval augmentation”, “materials science”, and specific task keywords (e.g. “crystal structure prediction”, “materials discovery”). Studies were included if they (1) applied LLMs or GPT-style models to a materials science problem, or (2) introduced methods for integrating materials domain knowledge (databases, literature, tools) with LLMs. After removing duplicates and screening titles and abstracts, a total of $\sim$21 peer-reviewed papers and preprints (2019–2025) were identified for detailed review. For each, we analyzed the problem addressed, the LLM or augmentation technique used, and key findings. The review is organized by application area. 

\subsection{Theoretical Background: LLMs and Retrieval-Augmented Generation}
\label{sec:background}
LLMs are neural networks (typically Transformer-based \citep{Vaswani2017}) trained to predict text, which endows them with a broad but implicit knowledge of language and facts from their training data. In practice, state-of-the-art LLMs like GPT-4 can answer questions, generate 
explanations, and perform reasoning tasks \citep{Brown2020-GPT3}. However, their knowledge is limited by the static training corpus and they may not know specialized scientific information not present in that data. Furthermore, LLMs have no direct access to tools or databases at runtime and can confidently output incorrect information. \textit{Retrieval-augmented generation} (RAG) addresses these limitations by augmenting the LLM with a retrieval mechanism \citep{Lewis2020-RAG}.

RAG is a framework that combines a neural text \textbf{retrieval} module with a text \textbf{generation} module to improve the quality of generated responses in knowledge-intensive tasks \citep{Oche2025ASR}. Formally, a RAG model augments a sequence-to-sequence (seq2seq) generator with access to an external text corpus (non-parametric memory) via a retriever \citep{Lewis2020, Karpukhin2020}. Given an input query $x$, the retriever $R$ selects a small subset of relevant documents $Z = \{z_1, z_2, \dots, z_K\}$ from a large corpus $\mathcal{C}$ (with $K \ll |\mathcal{C}|$) \citep{Karpukhin2020}. The generator then conditions on both the query $x$ and the retrieved documents $Z$ to produce an output $y$ (such as an answer or a descriptive text). Formally, the RAG model can be viewed as a latent variable generative model that defines a probability distribution over outputs $y$ by marginalizing over the retrieved documents $z_i$:
\begin{equation}\label{eq:rag-marginal}
    P(y \mid x) \;=\; \sum_{i=1}^{K} P_{\text{ret}}(z_i \mid x)\; P_{\text{gen}}(y \mid x, z_i)\,,
\end{equation}

\end{multicols}
\begin{figure}[h]
    \centering
    \includegraphics[width=0.65\textwidth]{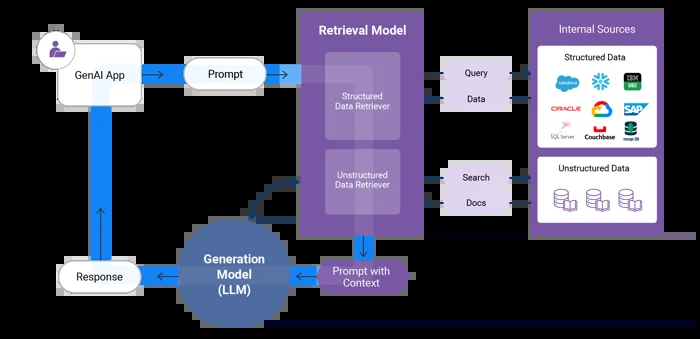}
    \caption{Illustration of a RAG Architecture.}
    \label{fig:rag_pipeline}
\end{figure}
\begin{multicols}{2}

where $P_{\text{ret}}(z_i \mid x)$ is the probability of retrieving document $z_i$ given query $x$ (the retriever's output distribution), and $P_{\text{gen}}(y \mid x, z_i)$ is the generator's conditional probability of producing $y$ given $x$ and a particular retrieved document $z_i$. In practice, $P_{\text{ret}}(z_i \mid x)$ is typically non-zero only for the top-$K$ retrieved items, providing a tractable approximation to the full sum over the corpus \citep{Lewis2020}. The retriever $R$ itself can be defined as a function $R(x, \mathcal{C}) \to Z$ that takes a query and returns a small subset $Z$ of corpus $\mathcal{C}$ (with $|Z|=K \ll |\mathcal{C}|$) likely to contain information relevant to $x$ \citep{Karpukhin2020}. By design, RAG models maintain two kinds of \textbf{memory}: a \emph{parametric memory} (the knowledge encoded in the generator’s weights) and a \emph{non-parametric memory} (the external text corpus accessed via retrieval) \citep{Lewis2020}. A standard RAG architecture is illustrated in Figure~\ref{fig:rag_pipeline} below. 
A key distinction between RAG and pure large language model (LLM) generation is the use of this external non-parametric knowledge source at inference time. Traditional LLM-based generation relies solely on the model’s internal parameters for knowledge, which can lead to \textbf{hallucinations} and factual inaccuracies when the model’s training data does not adequately cover the query’s topic \citep{Lewis2020}. In contrast, RAG explicitly grounds the generation of retrieved documents that serve as up-to-date evidence, enabling the model to generate content supported by those documents. This retrieval step means that RAG’s outputs can be more accurate and \textbf{factually correct} compared to generation from a standalone LLM, especially for knowledge-intensive queries. Empirically, \citep{Lewis2020} demonstrate that a RAG model generates more specific and factual responses than a parametric-only generator, since the retrieved text provides verified information that the generator can incorporate. Another benefit is that the knowledge in a RAG system can be easily \textbf{updated} by modifying the document index (or corpus) without retraining the generator, addressing the stiffness of LLMs that have fixed knowledge up to their training cutoff date. In summary, RAG introduces a modular architecture where a retrieval component supplies relevant context “just in time” for the generator, marrying the strengths of Information Retrieval (IR) with those of large-scale generation.

\section{Review of Key Existing Work}
\label{sec:review}
We now review how LLMs and RAG have been applied in several key areas of materials science. These applications illustrate the diverse ways in which combining language models with materials data and domain knowledge can assist research workflows. Summary of key work is in \cref{tab:rag_summary}.

\subsection{Crystal Structure Prediction}
Predicting the crystal structure of a material from its chemical composition is a long-standing challenge. Traditional crystal structure prediction methods (e.g. evolutionary algorithms or particle-swarm optimization) are computationally intensive. Recently, researchers have explored LLM-based generative approaches to propose likely crystal structures as a starting point for more refined calculations. For example, Antunes \textit{et al.} \citep{Antunes2024-CrystaLLM} introduced \textbf{CrystaLLM}, an autoregressive LLM trained on millions of crystallographic information files (CIFs) to generate plausible crystal structures in text form. CrystaLLM demonstrated the ability to produce realistic crystal structures for inorganic compounds not seen in training, which were validated via \textit{ab initio} calculations \citep{Antunes2024-CrystaLLM}. In a related effort, Gruver \textit{et al.} \citep{Gruver2024-InorganicGen} fine-tuned a GPT-style model on a large dataset of composition-structure-text triples and showed it could generate candidate inorganic materials that are both novel and predicted to be thermodynamically stable . These works indicate that LLMs can learn the complex structural chemistry encoded in databases like the Materials Project, offering a new route to \textit{de novo} crystal structure prediction. While still early-stage, LLM-generated structures may serve as creative hypotheses or seeds for more rigorous structure search algorithms. 

\subsection{Defect Analysis}
Crystalline defects (e.g. vacancies, dislocations, grain boundaries) critically influence materials properties. Analyzing and characterizing such defects often involves interpreting complex data from microscopy or simulations. LLMs augmented with retrieval and tools have shown potential in assisting defect analysis. One example is the \textbf{AtomAgents} system by Ghafarollahi and Buehler, which uses multiple AI agents (including an LLM coupled to materials databases and simulation tools) to design novel alloys \citep{Ghafarollahi2024-AtomAgents}. In their workflow, the LLM agent retrieves material data and even integrates physics-based simulation results; notably, the system demonstrated human-level proficiency in tasks like calculating alloy properties and analyzing defect structures in candidate materials \citep{Ghafarollahi2024-AtomAgents}. This suggests that an LLM can coordinate defect analysis by querying databases for known defect formation energies or by controlling simulation software to compute properties of defective structures. In another vein, vision-enabled LLMs (such as GPT-4V) have been used to examine microscopy images of materials for defect identification, combining textual context (e.g. captions, experiment details) with visual data. Early studies report that such multi-modal LLMs can classify and describe defects in micrographs with promising accuracy, effectively mimicking a materials scientist examining the images. Although dedicated computer vision models exist for defect detection, the flexibility of an LLM to integrate descriptive captions and background knowledge offers a unique advantage in generating human-readable analyses of defects. 

\subsection{Materials Search and Optimization}
One of the most ambitious goals is to harness LLMs as autonomous agents for discovering new materials and optimizing processes. Several recent systems have integrated LLMs with materials databases and simulation capabilities to propose novel compounds or formulations meeting target criteria. \textbf{ChatMOF} is a notable example: Kang \textit{et al.} \citep{Kang2024-ChatMOF} developed an AI assistant for metal–organic framework (MOF) design using GPT-4 as the engine. ChatMOF can retrieve information on known MOFs, predict properties, and even suggest new MOF structures with desired properties such as high surface area or specific gas adsorption performance \citep{Kang2024-ChatMOF}. Impressively, in evaluations ChatMOF achieved over 95\% accuracy on property predictions and was able to generate valid MOF structures that satisfied user-defined targets. Another system, \textbf{MatExpert} by Ding \textit{et al.} \citep{Ding2024-MatExpert}, attempts to decompose the human materials discovery process and emulate it with an LLM-based agent. MatExpert uses RAG to fetch relevant scientific knowledge and then proposes materials candidates, demonstrating success in tasks like identifying superconductors and catalysts by mimicking expert reasoning. In the field of batteries, Zhao \textit{et al.} \citep{Zhao2024-BatteryGPT} introduced a domain-specific model nicknamed \textbf{BatteryGPT} that leverages LLMs to extract and organize knowledge for battery material innovation. BatteryGPT was able to take user queries (e.g. for a type of electrode material with a certain voltage) and sift through databases and literature to recommend promising candidates, effectively transforming natural language inputs into experimental insights \citep{Zhao2024-BatteryGPT}. These examples illustrate that LLMs, when endowed with retrieval of materials data and possibly coupled with simulation tools, can serve as \textit{co-pilots} in materials discovery. They can rapidly generate hypotheses—such as suggesting new alloy compositions or crystal structures—by aggregating knowledge from vast sources, something human researchers would do much more slowly. At the same time, ensuring the scientific validity of LLM-generated suggestions remains a challenge; current systems incorporate verification steps (e.g. DFT calculations) to validate AI-proposed materials.

\subsection{Literature Mining}
The volume of materials science literature is enormous, making it difficult for researchers to stay up-to-date. LLMs have begun to play a role in mining this literature for relevant information and insights. Unlike earlier text-mining approaches that relied on keyword matching or static embeddings \citep{Tshitoyan2019}, modern LLMs can read passages and answer questions about them or summarize findings across papers. One application is the automated extraction of structured data from publications. For instance, Zhang \textit{et al.} \citep{Zhang2024-OFET} developed an LLM-driven agent for organic semiconductor research that reads papers to extract experimental parameters and device performance data for organic field-effect transistors (OFETs). This agent, built on GPT-4, achieved over 90\% accuracy in identifying key parameters from text and even suggested optimizations that improved device performance when tested in the lab \citep{Zhang2024-OFET}. Such capabilities hint at how LLMs could populate materials databases directly from literature, keeping them current. Another avenue is using LLMs to generate literature reviews or answer technical questions by synthesizing multiple sources.Zhang \textit{et al.} \citep{Zhang2024-Honeycomb} provides a question-answering system where an LLM is augmented with a curated materials science knowledge base and a retriever. It significantly outperformed vanilla LLMs on domain-specific QA benchmarks, demonstrating the value of augmenting the model with literature-derived knowledge. More experimentally, researchers have used GPT-4 to analyze scientific figures and captions in papers; for example, by feeding an LLM a paper's abstract and figure caption, it can deduce what material a micrograph image likely shows and the instrument used, thereby labeling datasets of images extracted from the literature. This combination of natural language understanding and vision could greatly accelerate the curation of research data from publications. Overall, LLMs are proving adept at \textit{literature mining} tasks—extracting, organizing, and summarizing information from the ever-growing body of scientific papers.

\subsection{Database Integration}
Materials science is rich in structured databases (for instance, crystallographic databases, materials property repositories, and synthesis knowledge bases). Integrating these data sources with LLMs enables more powerful query answering and analysis. RetrievM-augmented LLMs can directly draw on database entries to ground their responses. A prominent example is the \textbf{LLaMP} framework proposed by Chiang \textit{et al.} \citep{Chiang2024-LLaMP}. LLaMP employs a hierarchical agent approach where an LLM agent can query the Materials Project database in real-time and even invoke simulation tools to obtain high-fidelity data (e.g. retrieving a crystal structure and calculating its properties). By having access to a database of millions of materials, the LLM dramatically reduces hallucinations and can provide quantitatively accurate answers to questions about material properties \citep{Chiang2024-LLaMP}. For instance, if asked about the band gap of a given compound, the system will retrieve the experimental or calculated value from the database rather than relying on the LLM’s memory. The Materials Project API is just one example—similar integration has been done with other resources: ChatMOF, mentioned earlier, connects to a MOF-specific database; BatteryGPT was linked with battery materials data sources; and more generally, tool-using agents can run queries across multiple online databases. The key challenge in database integration lies in interfacing the structured query results with the LLM. Approaches like LLaMP use an intermediate reasoning layer (ReAct agents) to parse the user query, execute the correct API calls, and then feed the retrieved numeric or textual data back into the language model’s context. 

This approach has been shown to improve both the correctness and trustworthiness of LLM outputs for materials science questions, as the answers can be traced back to a database record. Database-integrated LLMs effectively function as intelligent front-ends, allowing researchers to ask complex questions in natural language (e.g. “Which known materials have a higher dielectric constant than X and are stable up to 1000~K?”) and receive answers grounded in the collective data of materials science.

\subsection{Multi-modal Retrieva-Augmented Generation}
Modern materials research often involves multimodal data—combinations of text, numeric data, chemical formulas, images (micrographs, spectra), etc. Extending RAG to handle multiple data modalities can further enhance LLM applications. One direction is incorporating visual data into the retrieval pipeline. For example, an LLM might retrieve not only text documents but also relevant figures or microscopy images. The advent of vision-capable LLMs (e.g. GPT-4V and other multi-modal transformers) means the model can directly analyze images alongside text. A recent system called \textbf{MatterChat} integrates a materials graph network with an LLM to create a multi-modal model that accepts both crystal structure data and textual context \citep{Tang2025-MatterChat}. By embedding 3D structural information (through a pre-trained interatomic potential model) into the LLM’s input, MatterChat was able to predict material properties and answer questions about materials with higher accuracy than text-only models \citep{Tang2025-MatterChat}. This showcases the benefit of giving the LLM direct access to structural data. Another example is the micrograph analysis pipeline described earlier: a two-step RAG approach where first text (captions/abstracts) is used to find candidate images, then a vision-LM examines those images to extract information (identifying the material or defects depicted). Multi-modal RAG can also mean linking experimental data (like spectra or phase diagrams) as part of the retrieval context for an LLM. By providing multiple modalities of evidence, the LLM can generate more comprehensive and accurate outputs—for instance, explaining an anomaly in a material’s properties by referencing both a microscopy image and relevant text from the experimental section of a paper. While still in nascent stages, these approaches hint at a future in which LLM-based agents seamlessly blend textual and visual (and even audio or other sensor) data in guiding materials research.

\end {multicols}
\begin{longtable}{@{}p{3.5cm}p{3.5cm}p{4.5cm}p{4.5cm}@{}}
\caption{Summary of LLM and RAG Applications in Materials Science} \label{tab:rag_summary} \\
\toprule
\textbf{Application Area} & \textbf{Approach/Tools} & \textbf{Key Contributions} & \textbf{Limitations} \\
\midrule
Crystal Structure Prediction & CrystaLLM, GPT-style generative models & Generated novel crystal structures validated via DFT & Needs verification of generated structures \\
\midrule
Defect Analysis & AtomAgents, GPT-4V with microscopy data & Analyzed defects using simulation tools and image data & Accuracy depends on image quality, simulation setup \\
\midrule
Materials Discovery & ChatMOF, MatExpert, BatteryGPT & Proposed materials meeting target criteria via RAG & Scientific validity of generated ideas must be confirmed \\
\midrule
Literature Mining & GPT-4-based extraction tools, HoneyComb QA & Extracted data, answered technical questions from literature & Hallucination possible if retrieval fails \\
\midrule
Database Integration & LLaMP framework, Materials Project integration & Enabled real-time data-grounded QA, reduced hallucination & Complex implementation, requires API/data access \\
\midrule
Multi-modal RAG & MatterChat, GPT-4V for images/spectra & Integrated structural/visual data for property prediction & Still nascent; needs improved fusion of modalities \\
\bottomrule
\end{longtable}

\begin{multicols}{2}

\section{Discussion}
\label{sec:discussion}

\subsection{Current Limitations}
The body of work reviewed above demonstrates the considerable promise of LLMs and RAG in materials science. Across disparate tasks—structure generation, defect analysis, knowledge extraction, and more—common themes emerge. First, LLMs excel at interfacing with human knowledge representations: they understand natural language queries and can generate human-readable explanations, which lowers the barrier for researchers to interact with complex data and tools. By augmenting LLMs with retrieval, these systems become more than just generative text bots; they effectively function as \textit{knowledge agents} that can combine learned language patterns with explicit factual databases or computation results. This has enabled impressive achievements, such as suggesting new materials that meet design criteria or automatically extracting useful data from a mountain of papers. 

However, several challenges and limitations are evident. One major issue is the reliability of LLM outputs. While retrieval helps ground answers, LLMs can still produce incorrect or unsubstantiated claims, especially if the retrieved context is insufficient or the model misinterprets the query. Ensuring verifiability (e.g. by providing references to data sources) is not fully solved. In the materials domain, where mistakes can lead to costly experimental dead-ends, maintaining a high level of trust in AI-generated suggestions is crucial. Techniques like the uncertainty+confidence metrics proposed in LLaMP \citep{Chiang2024-LLaMP} and incorporating human expert feedback will be important for validation. Another limitation is that current systems tend to handle relatively well-defined tasks (like question answering, property prediction) but are less adept when open-ended creativity or deep reasoning is required. For instance, LLMs might suggest a new material composition but they cannot yet reliably predict unforeseen challenges in synthesizing that material. The scope of multi-step reasoning an LLM can perform is also constrained by context length and the model’s training. Multi-agent frameworks (as seen in AtomAgents, MatExpert) partially address this by breaking tasks into smaller LLM-invocations with planning, but this adds complexity and potential failure points.
There are also practical considerations. Implementing RAG pipelines in research settings requires technical expertise to connect LLMs with databases and simulation codes. The large computational cost of state-of-the-art LLMs (often accessible only via APIs or large servers) can be a barrier; some domain researchers may prefer smaller fine-tuned models for cost and privacy reasons. Moreover, as with any data-driven approach, biases in the training data can propagate. If the literature or databases used contain biases (e.g. over-reporting certain material types), the LLM might inadvertently amplify those in its outputs.

\subsection{Future Potentials and Emerging Directions}

The integration of RAG into materials science is still in its infancy, and many powerful extensions remain on the horizon. Here we outline several promising future directions, including autonomous research loops that use RAG, generative workflows for closed-loop discovery, and human-in-the-loop systems that capitalize on RAG’s strengths.

\subsubsection{Autonomous Research Loops with RAG:}

A bold vision for the future is the concept of autonomous laboratories or self-driving materials research loops, where AI agents design experiments, carry them out with robotic platforms, analyze results, and then decide on the next experiments \citep{Wang2024}. In such a setting, an LLM endowed with retrieval could function as the “brain” of the lab, dynamically gathering background knowledge and assimilating new data as experiments progress. Retrieval augmentation is critical here because the agent must remain grounded in reality: as soon as new experimental data are produced, they should be fed back into the LLM’s context (a form of retrieval from the lab’s database) so that subsequent decisions are based on up-to-date information. Likewise, when the agent encounters an unexpected result, it can retrieve relevant literature or prior data to hypothesize an explanation or adjust the experimental plan. For example, imagine an autonomous synthesis loop searching for a new superconductor: the LLM proposes a composition to test, the robotic lab synthesizes and measures it, and the measured Tc is far lower than predicted. The LLM agent could retrieve past studies of similar compounds where a certain structural distortion was responsible for Tc suppression, realize the same issue might be occurring, and therefore pivot the search in a new direction. Without RAG, the agent would lack this adaptability, as it wouldn’t have access to insights beyond its trained knowledge. We can envision one agent specialized in proposing candidates (drawing on generative modeling and prior data via retrieval), another in analyzing characterization data (using multi-modal RAG to interpret spectra/images against known references), and a top-level agent coordinating the loop. This kind of system could dramatically accelerate materials discovery, as it effectively runs 24/7, learning and adapting from its own results by constantly retrieving and ingesting new knowledge. Early steps are already visible in initiatives like the \textit{Materials Open Platform} and other AI-driven closed-loop experiments, but the added layer of LLM+RAG introduces a powerful generalist reasoning capability. In the coming years, demonstrating a self-driving lab that discovers a material completely autonomously with the help of an LLM orchestrator (using RAG to stay scientifically smart) would be a groundbreaking milestone.

\subsubsection{Generative Workflows and Closed-Loop Design:}

Generative models (whether graph-based, diffusion, or autoregressive) have become key tools for proposing new material candidates. RAG can be the connective tissue that embeds these models into closed-loop design workflows. In essence, RAG allows generative models and evaluators to talk to each other through the medium of an LLM. A case in point was MatAgent (discussed earlier), where an LLM brokered the interaction between a generator and a predictor by retrieving intermediate results and deciding how to refine proposals \citep{Takahara2025}. Generalizing this, we expect future AI-driven workflows to involve an LLM that can call on many specialized components (simulation codes, synthesis planning tools, etc.), using retrieval to supply each component with the information it needs. This is related to the idea of an \textit{AI chemist} that maintains a memory of everything tried so far and has access to all chemical rules. Such a system could, for example, generate a series of candidate polymers, simulate their dielectric constants via a quantum chemistry tool, retrieve the simulation results and prior known polymer data, then prompt itself to find patterns and suggest a next set of candidates focusing on the promising chemical motifs. The LLM’s role is to ensure the loop is informed by both the newly generated data and the established scientific knowledge at every step. 

One concrete benefit of RAG in these workflows is improved sample efficiency. Closed-loop optimization in materials often suffers from needing many cycles to converge (because the search space is huge). But if the AI can retrieve hints like “Increasing the Zn content tended to raise the band gap in prior experiments” or “A similar compound was unstable because of moisture sensitivity,” it can make more informed jumps in the search space, potentially cutting down the number of iterations. In practice, we already see simpler versions of this in literature: for instance, a recent study used an LLM to analyze DFT results and retrieve chemical insights to explain why certain generated molecules failed, then used that explanation to filter the next round of generation \citep{jia2024llmatdesign}. Extending that idea, the LLM could dynamically modify the objectives or constraints of the generative model based on retrieved domain knowledge (“according to the phase diagram retrieved, we should avoid compositions in this range, focus search where a second phase won’t form”). Thus, generative workflows guided by RAG could become much more efficient explorers of materials space.

Another emerging direction is the use of reinforcement learning (RL) where the reward signal is augmented by knowledge. For example, an LLM agent could generate candidate experiments and get a reward not only from the experimental result but also from a heuristic based on retrieved literature (e.g., penalizing proposals that violate known chemical stability rules). This hybrid of RL with knowledge augmentation may address the challenge of sparse or expensive real-world rewards by leveraging the “wisdom” encoded in databases as a shaping function. Overall, the future trend is toward LLM-centric systems that unify generation, evaluation, and knowledge integration in a continuous loop—bringing us closer to autonomous scientific discovery pipelines that iterate efficiently towards optimized materials.

\subsubsection{Human-in-the-Loop Design:}

\begin{figure}[htbp]
  \centering
  \includegraphics[width=0.85\linewidth]{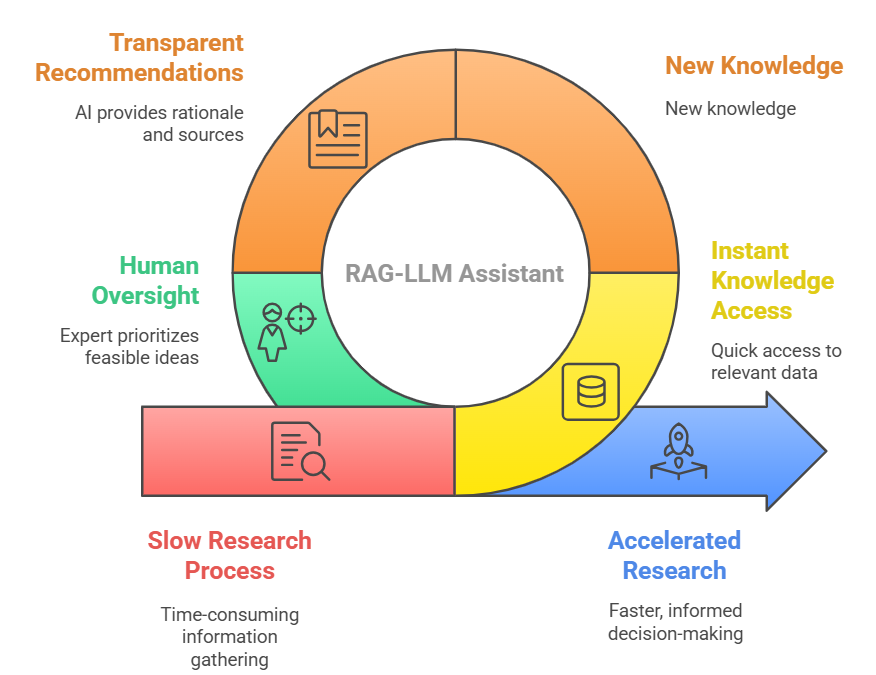}
  \caption{Design of Human in the loop integrated RAG/LLM systems.}
  \label{fig:human_in_the_loop}
\end{figure}

While full autonomy is a tantalizing goal, in the near term a more practical and immediately beneficial application of RAG is in human-in-the-loop design systems. Figure~\ref{fig:human_in_the_loop} illustrates a generalized example of human in the loop systems, integrated with RAG-LLM. In this paradigm, the RAG-empowered LLM acts as an assistant or co-pilot to the human researcher, enhancing their decision-making with instant access to knowledge and suggestions. The human still guides the overall goals, provides intuition, and makes final decisions, but the AI greatly amplifies the breadth and depth of information considered at each step. RAG is crucial in this context because trust and verification are paramount when an AI is advising scientific decisions. By retrieving sources and evidence for its statements, the LLM can provide the human designer with not just recommendations but also the rationale behind them. This turns the AI from a mysterious oracle into a transparent partner that cites chapter and verse, making it easier for the human to gauge whether a suggestion is credible.
\end{multicols}

\begin{multicols}{2}
We foresee tools where a materials scientist can query an interactive system in natural language throughout the R\&D process. For instance, during a brainstorming session, one might ask: "What known crystal structures might accommodate a $+2$ cation and show ferroelectric behavior?" The RAG-LLM could instantly retrieve relevant entries from a crystal structure database (e.g. perovskite oxides, fluorites, etc.) and respond: "Many perovskites with general formula ABO$_3$ exhibit ferroelectricity when A is $+2$ and B is $+4$ (e.g. BaTiO$_3$). You could consider structures like the perovskite family; indeed, BaTiO$_3$ and PbTiO$_3$ are classic ferroelectrics. Alternatively, some layered structures (Aurivillius phases) incorporate $+2$ cations and are ferroelectric. Such an answer not only gives the suggestion (perovskites, layered oxides) but backs it up with references and examples, saving the researcher enormous time scanning textbooks or memory. The human can then drill down: "What if I substitute a smaller cation at the A-site of BaTiO$_3$?" The LLM might retrieve studies on CaTiO$_3$ or SrTiO$_3$ and warn that those become paraelectric, thus guiding the user away from a known dead end.

Another aspect of human-in-the-loop systems is that they can leverage the creativity of AI while keeping a human filter to ensure feasibility and prioritize ideas. An LLM might generate 50 hypothetical new alloys with certain features, but rather than executing all of them, it presents them to the expert with literature support for each. The expert can quickly scan: "This one looks interesting and there’s a paper suggesting its microstructure would be stable – let’s try that first." In essence, RAG helps compress the information gathering and analysis phase that an expert would normally do (reading dozens of papers, recalling prior knowledge) into a digestible AI output, so the expert can focus on high-level strategic decisions and experimental execution. Early user studies in other domains have shown that such AI assistants can significantly enhance human problem-solving productivity, and similar gains are anticipated in materials engineering. 

\subsubsection{Domain-Specialized LLMs:} Thus far, many applications use general models (like GPT-3.5/4) with prompting or lightweight fine-tuning. An important next step is developing large models trained or adapted specifically on materials science knowledge (similar to how BioGPT was developed for biomedical text). Early efforts such as MatSciBERT and recent domain LLMs \citep{Xie2023-DARWIN} indicate this is feasible. A materials-optimized LLM could capture nuances of chemistry terminology and quantitative reasoning better, reducing reliance on prompts for basic domain facts.
\subsubsection{Better Integration of Knowledge and Physics:} Retrieval augmentation could be extended beyond pulling static data to integrating knowledge graphs of materials relationships or incorporating physical constraints. For example, an LLM that knows basic thermodynamic laws or crystal symmetry rules inherently would be less likely to propose impossible compounds. Hybrid models that combine LLMs with symbolic reasoning or physics-based modules are a promising avenue to bring scientific consistency to AI suggestions.

\subsubsection{Robust Multi-modal Capabilities:} As multi-modal RAG matures, future systems might allow a researcher to input a mix of data—e.g. a phase diagram image, a target property, and some text notes—and get coherent analysis or recommendations. Achieving seamless understanding across text, images, and potentially spectra or simulation outputs will likely require new model architectures or training paradigms, but could greatly accelerate tasks like diagnosing material failures or screening candidates.

\section{Conclusion}
In this review, we systematically examined the role of large language models (LLMs) and retrieval-augmented generation (RAG) in accelerating crystalline material discovery. Our analysis covered key application areas including crystal structure prediction, defect analysis, materials search and optimization, literature mining, database integration, and multi-modal retrieval. The integration of LLMs with domain-specific retrieval modules has demonstrated significant potential in reducing the vast material search space, improving accuracy and interpretability, and assisting researchers in navigating complex data and literature.

However, several challenges remain, including the reliability and verifiability of LLM-generated outputs, integration complexity, and computational costs. The susceptibility of LLMs to hallucination underscores the necessity of retrieval augmentation to ground their predictions in verified external knowledge sources. We identified promising future directions such as autonomous research loops, generative workflows, human-in-the-loop design systems, domain-specialized LLMs, enhanced integration of physical laws and scientific knowledge, and robust multi-modal capabilities.

Ultimately, leveraging RAG frameworks to embed LLMs into materials research processes presents an exciting frontier. These developments hold the promise of transforming conventional materials discovery workflows, accelerating innovation, and significantly advancing technological capabilities across electronics, optics, biomedical applications, and energy storage domains.

\section*{Acknowledgments}
This work (J.A) was supported by A.B's startup funding. The authors acknowledge the use of facilities and instrumentation at the UT Knoxville Institute for Advanced Materials and Manufacturing (IAMM) supported in part by the National Science Foundation Materials Research Science and Engineering Center program through the UT Knoxville Center for Advanced Materials and Manufacturing (DMR-2309083). We extend our gratitude to Dr. Yishu Wang for their invaluable feedback.

\newpage
\bibliographystyle{plain}
\bibliography{references}

\fussy 
\end{multicols}

\section*{Author Contributions} Conceptualization: J.O.A., A.B.; Methodology (review scope, inclusion/exclusion criteria): J.O.A., A.B.; Literature search \& data curation: J.O.A.; Formal analysis/synthesis: J.O.A.; Visualization: J.O.A.; Writing—original draft: J.O.A.; Writing—review \& editing: J.O.A., A.B.; Supervision: A.B.; Project administration: A.B.; Resources \& funding acquisition: A.B.

\section*{Conflict of Interest}
The authors confirm there is no conflict of interest.

\end{document}